%% file: paper.tex
\documentclass{article}

\usepackage{arxiv}

\usepackage[utf8]{inputenc} 
\usepackage[T1]{fontenc}    
\usepackage{hyperref}       
\usepackage{url}            
\usepackage{booktabs}       
\usepackage{amsfonts}       
\usepackage{nicefrac}       
\usepackage{microtype}      
\usepackage{lipsum}
\usepackage{subcaption}
\usepackage{graphicx}
\usepackage[symbol]{footmisc}

\graphicspath{ {./images/} }

\usepackage{amsbsy}
\usepackage{amsmath}

\usepackage[numbers]{natbib}

\title{MARS-Gym: A Gym framework to model, train, and evaluate Recommender Systems for Marketplaces}

\author{
  Marlesson R. O. Santana \footnote[1]{Authors contributed equally to this research.} \\
  Deep Learning Brazil\\
  Federal University of Goiás, Brazil\\
  \texttt{marlessonsa@gmail.com} \\
   \And
  Luckeciano C. Melo \footnote[1]{Authors contributed equally to this research.} \\
  Deep Learning Brazil\\
  Federal University of Goiás, Brazil\\
  \texttt{luckeciano@gmail.com} \\
   \And
  Fernando H. F. Camargo \footnote[1]{Authors contributed equally to this research.} \\
  Deep Learning Brazil\\
  Federal University of Goiás, Brazil\\
  \texttt{fernando.camargo.ai@gmail.com} \\
   \And
  Bruno Brandão \footnote[1]{Authors contributed equally to this research.} \\
  Deep Learning Brazil\\
  Federal University of Goiás, Brazil\\
  \texttt{brunobrandao1523@gmail.com} \\
   \And
  Anderson Soares \\
  Deep Learning Brazil\\
  Federal University of Goiás, Brazil\\
  \texttt{anderson@inf.ufg.br} \\  
   \And
  Renan M. Oliveira \\
  iFood Research\\
  \texttt{renan.oliveira@ifood.com.br} \\  
   \And
  Sandor Caetano \\
  iFood Research\\
  \texttt{sandor.caetano@ifood.com.br} \\    
}

\begin{document}
\maketitle

\footnotetext[1]{Authors contributed equally to this research.}

\input{tex/abstract}

\keywords{marketplace recommendation \and reinforcement learning \and off-policy evaluation \and fairness}

\input{tex/intro}
\input{tex/related-work}
\input{tex/preliminaries}
\input{tex/framework}

\input{tex/experiments}
\input{tex/conclusions}
\input{tex/acknowledgments}

\bibliographystyle{plainnat}  
\bibliography{references}

\newpage
\appendix

\end{document}

%% file: tex/abstract.tex
\begin{abstract}
  Recommender Systems are especially challenging for marketplaces since they must maximize user satisfaction while maintaining the healthiness and fairness of such ecosystems. In this context, we observed a lack of resources to design, train, and evaluate agents that learn by interacting within these environments. For this matter, we propose \textit{MARS-Gym}, an open-source framework to empower researchers and engineers to quickly build and evaluate Reinforcement Learning agents for recommendations in marketplaces. \textit{MARS-Gym} addresses the whole development pipeline: data processing, model design and optimization, and multi-sided evaluation. We also provide the implementation of a diverse set of baseline agents, with a metrics-driven analysis of them in the Trivago marketplace dataset, to illustrate how to conduct a holistic assessment using the available metrics of recommendation, off-policy estimation, and fairness. With \textit{MARS-Gym}, we expect to bridge the gap between academic research and production systems, as well as to facilitate the design of new algorithms and applications.
\end{abstract}

%% file: tex/intro.tex
\section{Introduction}

Recommender Systems (RS) are essential tool to offering a high-quality experience to users on online platforms. These systems become even more critical for marketplaces. In this specific scenario, recommenders should also consider all partners, rather than assume a user-centric perspective. This requirement is critical to the healthiness and fairness of such ecosystems. Despite the recent progress in building successful recommendation models to real-world applications \cite{amatriain2015recommender,logesh2019exploring}, and, in special, for marketplaces \cite{ubereats, bart}, we observed a lack of resources to experiment models that learn through interaction within these ecosystems effectively. To the best of our knowledge, there are no frameworks that help researchers and engineers to quickly design, train, and evaluate Reinforcement Learning (RL) agents to optimize the dynamics of marketplaces, with special focus on real-world applications.

In this work, we present \textit{MARS-Gym} (\textbf{MA}rketplace \textbf{R}ecommender \textbf{S}ystems Gym), a benchmark framework for modeling, training, and evaluating RL-based recommender systems for marketplaces. Three main components composes the framework. The first one is a highly customizable module where the consumer can ingest and process a massive amount of data for learning using spark jobs. The second component was designed for training purposes. It holds an extensible module built on top of PyTorch \cite{paszke2019pytorch} to design learning architectures. It also possesses an OpenAI's Gym \cite{Brockman2016OpenAI} environment that ingests the processed dataset to run a multi-agent system that simulates the targeted marketplace. 

Finally, the last component is an evaluation module that provides a set of distinct perspectives on the agent's performance. It presents not only traditional recommendation metrics but also off-policy evaluation metrics, to account for the bias induced from the historical data representation of marketplace dynamics. Finally, it also presents fairness indicators to analyze the long-term impact of such recommenders in the ecosystem concerning sensitive attributes. This component is powered by a user-friendly interface to facilitate the analysis and comparison between agents.

In summary, our key contributions are the following:

\begin{itemize}
    \item An open-source framework to empower researchers and engineers to build, train, and evaluate RL agents for marketplace recommendation;
    \item The implementation of a diverse set of contextual bandits \cite{DBLP:journals/corr/abs-1811-04383} to serve as baselines and illustrations of how \textit{MARS-Gym} works; and
    \item A deep, metrics-driven analysis on the performance of such bandits, with special focus on the interpretation of off-policy metrics and sources of unfairness for sensitive attributes.
\end{itemize}

This paper is organized as follows. In Section \ref{sec:related_work}, we describe related work. Section \ref{sec:preliminaries} presents the theoretical background. Section \ref{sec:framework} details the framework usage, its internal components, and the evaluation protocol. Section \ref{sec:experiments} presents the experimental results and analysis for the baselines agents. Finally, Section \ref{sec:conclusions} concludes and shares our thoughts for future work.

%% file: tex/related-work.tex
\section{Related Work}\label{sec:related_work}

We identified some works which focused on marketplace recommendations. \citeauthor{ubereats} \cite{ubereats} proposed a model based on multi-armed bandits and UCB (Upper Confidence Bound) exploration. It was trained via multi-objective optimization to address exploration and diversity of recommendations in a food delivery marketplace. \citeauthor{bart} \cite{bart}, on the other side, formulated the problem as a contextual bandit in an off-policy training scenario to recommend not only items but also an associated explanation in an online music streaming service. For comparison, our work also addresses marketplace recommendation with sequential decision-making formulation; however, we propose a framework and benchmark environment to improve research and development in this subject, rather than new models and applications.



In terms of environments for RS evaluation, we identified works that focused on open-sourcing datasets \cite{Bennett07thenetflix, 10.1145/2827872, 7410843}, elaborating competitions \cite{Knees_etal:RSC:2019, DBLP:journals/corr/abs-1810-01520, 10.1145/3109859.3109954}, and releasing environments or platforms to engage researchers. In the last line of contribution, which is closely related to our work, we highlight \textit{RecoGym} \cite{rohde2018recogym} and \textit{PyRecGym} \cite{10.1145/3298689.3346981} platforms. Like our work, they also created OpenAI's Gym environments for sequential interaction, which enables the development of RL agents in RS.
However, those platforms assumed a user-centric perspective where the episodes are users' sessions, and the objective is to maximize their satisfaction. On the other side, our work assumes the perspective of the whole marketplace, simulating several users acting in each episode, aiming to maximize the satisfaction of all sides involved (users, suppliers, third parties) and, therefore, the healthiness and fairness of this ecosystem.

A Recent work proposed \textit{RecSim} \cite{ie2019recsim}, a platform for authoring environments for RS. Like our work, it also enables RL agents via configurable simulations. Despite that, \textit{RecSim} requires the implementation of some layers of abstraction to emulate specific aspects of user behavior, focusing on challenging fundamental assumptions of RL models from the perspective of RS. On the other hand, \textit{MARS-Gym} is data-driven by ingesting datasets, and aims to facilitate the research and development of applications for marketplace recommendation.

\textit{MARS-Gym} has a special focus on multi-sided evaluation. In terms of off-policy metrics and counterfactual estimation, many works propose new indicators \cite{10.1145/3159652.3159687, 10.5555/3104482.3104620}. We also highlight a recent body of applications, such as search engines \cite{10.1145/2740908.2742562}, slate recommendation \cite{NIPS2017_6954}, and even marketplaces \cite{10.1145/3269206.3272027}. The current work implements such metrics to compare against traditional recommendation metrics, as well as to diagnose any source of bias induced by the historical dataset.

Finally, in terms of fairness, there is a large body of recent work that formulates this concept for marketplaces \cite{DBLP:journals/corr/Burke17aa, 10.1145/3079628.3079657} and use it as evaluation protocol or even as optimization objective \cite{10.1145/3269206.3272027, patro2019incremental,ubereats}. Our work presents fairness metrics either to ensure that models account for the satisfaction and visibility for all involved sides and also to diagnose any source of bias in sensitive attributes.

%% file: tex/preliminaries.tex
\section{Preliminaries}\label{sec:preliminaries}

In this section, we introduce the notation and formalize the idea of Reinforcement Learning for Recommender Systems, the off-policy paradigm for evaluation and learning, and the notions of fairness applied in this work.

\subsection{Reinforcement Learning and Recommender Systems}


We formalize a recommendation system by the main aspects of its composition. We call a policy $\pi$, the primary means of making decisions. Thus, in the case where a dataset $\mathcal{D}$ is available, the policy used to acquire it is called a collection policy $\pi_c$. In order to make recommendations, the policy takes into account $x \in  \mathcal{X}: \mathbb{U} \times \mathcal{C}$ composed of the current user $u \in \mathbb{U}$ and contextual information $c \in \mathcal{C}$. Finally, the policy selects an action $a \in \mathcal{A}$ -- according to $\pi(a \mid x)$ -- as a recommendation to the user. 

Traditional approaches for Recommender Systems consider this problem in a supervised learning setting, where the model is trained by computing the error between its current response and the optimal answer.  However, by using this approach, we not only assume the recommendation as a stationary problem but also ignores the incomplete view from users' interests, once that the feedback is only partially observable. In order to take advantage of those scenarios and effectively learn from them, we decided to use Reinforcement Learning.



\begin{figure}[h]
  \centering
  \includegraphics[width=0.6\linewidth]{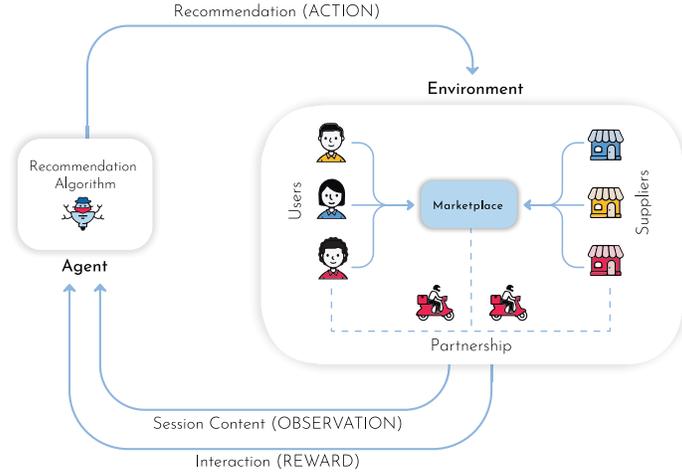}
  \caption{The interaction between the Agent and the Environment as a Markov Decision Process (MDP).}
  \label{fig:gym}
\end{figure}

Reinforcement Learning is a machine learning technique based on trial, error, and feedback. It learns on data generated by interacting with an environment. The RL problem formulation is formalized as a Markov Decision Process (MDP) \cite{Sutton2018}, illustrated in Figure \ref{fig:gym}. In an MDP, an \textit{agent} is whatever form that makes decisions; at a time step $t$, it receives from the environment \textit{state} $s_{t} \in \mathcal{S}$ and chooses from a finite set of \textit{actions} $a_{t} \in \mathcal{A}$. The environment changes, following a dynamics $p : \mathcal{S} \times \mathcal{S} \times \mathcal{A} \rightarrow [0, \infty)$, which represents the probability density of the next state. Finally, the environment sends a feedback \textit{reward} $r : \mathcal{S} \times \mathcal{A} \rightarrow [0, 1]$. The ultimate objective of an RL agent is to maximize the cumulative reward, i.e., $\max \mathbb{E} [\sum_{t=0}^{T} \gamma^{t} r (s_{t}, a_{t})]$, where $\gamma$ is a discount factor to account for delayed rewards. On the other side, Contextual Bandits simplifies this formulation by considering the problem with immediate rewards. Therefore, we do not need to consider this delay.
\subsection{Off-Policy Evaluation and Learning}\label{sec:off-policy}



 \textit{MARS-Gym} addresses the process of training a Recommender System as a off-policy, batch learning from partially observed feedback in a historical dataset. This consideration is advantageous because we do not require any additional interaction with production systems, which is often very costly. This advantage is valid for learning and evaluation, which makes iteration cycles faster and helps to detect poor policies before any deployment.
 
 Nevertheless, this is acknowledged as a hard problem, because such data is biased towards the collection policy (predictions favored by the historical algorithm will
be over-represented) and is incomplete (feedback for other predictions will not be available) \cite{10.5555/3045118.3045206}. To address this problem, we need counterfactual estimators \cite{JMLR:v14:bottou13a}, in order to estimate how other systems would have performed if they had been recommending items in the place of the collection policy. Several works applied these estimators for both off-policy evaluation \cite{10.1145/2740908.2742562, 10.5555/3104482.3104620} and learning \cite{bart, 10.5555/3045118.3045206}.

For evaluation, \textit{MARS-Gym} implements three main estimators. The first one is called the Direct Method (DM), which forms an estimate $\hat{\varrho}(x, a)$ of the expected reward conditioned on the context and action. We estimate the policy value by using the following Monte-Carlo estimator (Equation \ref{eq:dm}) over the historical dataset.

\begin{equation}\label{eq:dm}
    \hat{v}_{DM}^{\pi_{e}} = \frac{1}{\lvert \mathcal{D} \rvert}\sum_{x \in \mathcal{D}} \sum_{a \in \mathcal{A}} \pi_{e}(a \mid x) \hat{\varrho}(x, a), 
\end{equation}

\noindent where ${\varrho}(x, a) = \mathbb{E}_{(x, a, r) \sim \pi_{c}} [r | x, a]$. The advantage of this estimator is because it is simple and has low variance. However, it does not take into consideration the distribution from the evaluated policy, and its approximation will directly depend on the collection policy data, which is often biased.

The second estimator is called Inverse Propensity Score (IPS), which forms an estimate of the collection policy $\pi_{c}$ and uses Importance Sampling to re-weight rewards generated by it such that they are unbiased estimates of the evaluated policy $\pi_{e}$. Hence, the policy value is estimated by Equation \ref{eq:ips}. While the IPS estimator is less prone to induce bias when compared to DM, the former has a much larger variance, especially when our estimation of the collection policy is small for the context and action evaluated. There are heuristics to control this problem by introducing a bias-variance trade-off. They consist of capping and normalizing importance weights \cite{10.5555/3104482.3104620}.

\begin{equation}\label{eq:ips}
    \hat{v}_{IPS}^{\pi} = \frac{1}{\lvert \mathcal{D} \rvert}\sum_{(x, a, r) \in \mathcal{D}} \frac{\pi_{e} (a \mid x)}{\hat{\pi_{c}}(a \mid x)} r.
\end{equation}

The third estimator is the Doubly Robust (DR) Estimator in equation \ref{eq:dr}, which combines both reward and collection policy estimates, using the former as a baseline and the latter for correction. Hence, if at least one of them is accurate, then DR is also accurate. We will guide our analysis mainly using this metric.

\begin{equation}\label{eq:dr}
    \hat{v}_{DR}^{\pi} = \frac{1}{\lvert \mathcal{D} \rvert}\sum_{(x, a, r) \in \mathcal{D}}
    \bigg[ \frac{\pi_{e} (a \mid x)}{\hat{\pi_{c}}(a \mid x)} (r - \hat{\varrho}(x, a)) + 
    \sum_{a \in \mathcal{A}} \pi_{e}(a \mid x) \hat{\varrho}(x, a) \bigg].
\end{equation}

Finally, for learning, we apply the Counterfactual Risk Minimization (CRM) \cite{10.5555/3045118.3045206}. This learning principle uses one of the counterfactual estimators to modify the log-likelihood. Instead of evaluation policy, we consider an uniform distribution over actions $\mathcal{U}(a)$, to account for the fact that the data come from the production policy $\pi_{c}$ and not from an uniform random experiment \cite{bart}. Concretely, the current version of \textit{MARS-Gym} implements CRM via IPS, as shown in Equation \ref{eq:crm}, where $p_{\boldsymbol{\theta}}$ is a distribution, parameterized by $\boldsymbol{\theta}$, from which we derive the agent policy $\pi$.

\begin{equation}\label{eq:crm}
    \mathcal{L}(\boldsymbol{\theta}) = - \frac{1}{\lvert \mathcal{D} \rvert}\sum_{(x, a, r) \in \mathcal{D}} \frac{\mathcal{U}(a)}{\hat{\pi_{c}}(a \mid x)} \log {p_{\boldsymbol{\theta}}}(r \mid x, a).
\end{equation}

\subsection{Fairness Evaluation}\label{sec:fairness}

We commonly conceive Recommender Systems training as data-driven, optimization problems. Due to this reason, they are very prone to reproduce dataset biases as well as to disregard societal and economic impacts when they solely maximize user satisfaction. More than that: as automated decision-making systems, they are capable of \textit{changing} the data distribution, i.e., the way people interact within the marketplace. Therefore, we need to handle the long-term impact of such systems in production \cite{10.1145/3351095.3372878}.

For this, we must evaluate RS by using metrics whose purpose is to diagnose sources of unfairness that could impact the entities in the marketplace. In the context of recommendation, we focus on two main sources. First, when the user has sensitive attributes that could potentially cause any form of disadvantage or poorer experience for them in the platform. Second, when partners do not have fair opportunity to be presented to users: partner diversity and visibility are intrinsically related to the healthiness of marketplaces, especially because they are acknowledged to suffer from ``superstar economics" \cite{RePEc:aea:aecrev:v:71:y:1981:i:5:p:845-58}.

In \textit{MARS-Gym}, we consider the notion of fairness in three perspectives \cite{10.1145/3038912.3052660}. First, on the perspective of \textit{disparate treatment}, which arises when a decision-making system provides different outputs for groups of users with the same or similar non-sensitive attributes but different values of sensitive attributes. Mathematically, to avoid it, we need to ensure Equation \ref{eq:treatment},  where $z$ is the sensitive attribute in the analysis:

\begin{equation}\label{eq:treatment}
    \pi(a \mid x) = \pi(a \mid x, z).
\end{equation}

The second perspective is about \textit{disparate impact}, which happens when the decision outputs disproportionately benefit or hurt a group of users who share a particular sensitive attribute value. To avoid it, we need to guarantee Equation \ref{eq:impact},  where $\mathcal{Z}$ represents the set with all possible values of a sensitive attribute $z$:

\begin{equation}\label{eq:impact}
    \pi(a = a_{k}\mid z_{i}) = \pi(a = a_{k} \mid z_{j}), \\
    \forall a_{k} \in \mathcal{A}, \nonumber
    \forall z_{i}, z_{j} \in \mathcal{Z}. \nonumber
\end{equation}

Finally, we also consider the perspective of \textit{disparate mistreatment}. It arises when the \textit{misclassification rates} differ for groups of users having different values for a given sensitive attribute. To compute these rates, we need ground-truth actions for each interaction. In the current version of the \textit{MARS-Gym}, we measure it by using true positive rates. Therefore, we need to satisfy Equation \ref{eq:mistreatment}, where $a^{*}$ is the ground-truth action:

\begin{equation}\label{eq:mistreatment}
    \pi(a = a^{*}\mid z_{i}, a^{*} = a_{k}) = \pi(a = a^{*} \mid z_{j}, a^{*} = a_{k}), \\
    \forall a_{k} \in \mathcal{A}, \nonumber
    \forall z_{i}, z_{j} \in \mathcal{Z}. \nonumber
\end{equation}

In the following section, we describe how \textit{MARS-Gym} verifies these mathematical requirements in the evaluation module.

%% file: tex/framework.tex
\section{\textit{MARS-Gym} Architecture}\label{sec:framework}

In this section, we present how MARS-Gym simulates the marketplaces, as well as the design of internal modules that provides features to model, train, and evaluate Recommender Systems. The source code of \textit{MARS-Gym} is available online in \url{https://github.com/deeplearningbrasil/mars-gym}.

\subsection{MDP Simulation}\label{sec:mdp_simulation}

The main \textit{MARS-Gym} feature is the dynamic modeling of a marketplace as an MDP. The framework ingests a \textit{ground-truth dataset} from the real marketplace in order to perform a realistic simulation. Then, it generates an OpenAI's Gym environment where the data drives the internal transitions resulted from the interaction within the recommender. The consumer of \textit{MARS-Gym} must provide the ground-truth dataset. 

We provide a protocol with the requirements that the data should follow to build an environment, with two main requirements. First, a list of interactions between users and partners, with all contextual information and a binary variable that explicit whether each interaction was successful (e.g., a buy or click out). Second, the metadata that describes users and partners.

\begin{figure}[h]
  \centering
\includegraphics[width=\linewidth]{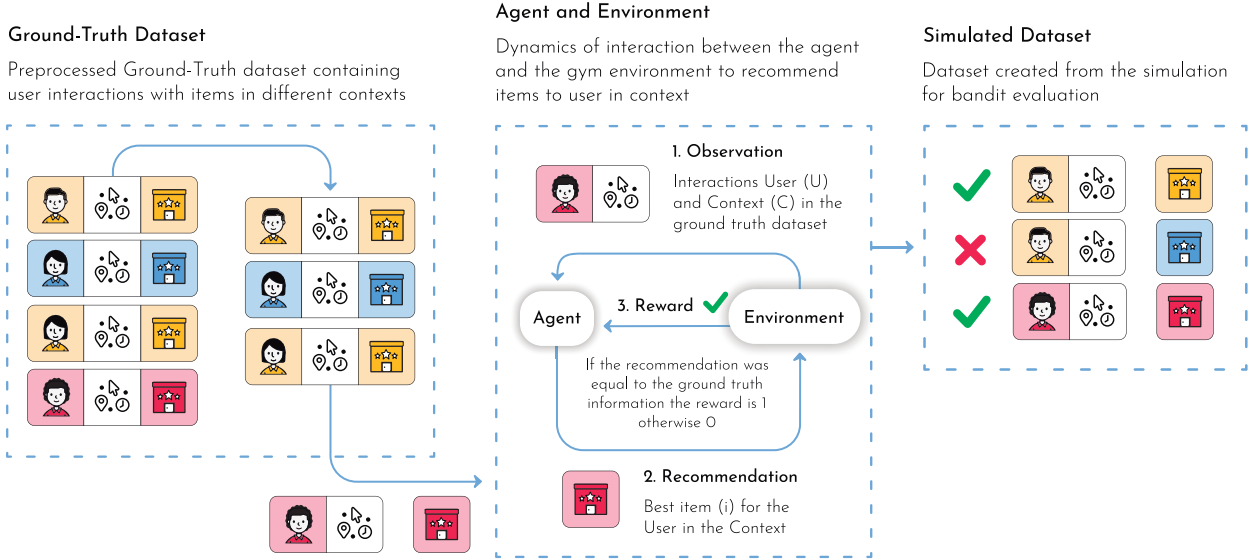}
  \caption{Diagram flow of \textit{MARS-Gym}, from dataset ingestion for environment generation to the MDP simulation.}
  \label{fig:process}
\end{figure}

Figure \ref{fig:process} presents how \textit{MARS-Gym} simulates the dynamics of the marketplace. First of all, the framework filters only successful interactions, once they are the available source of the true reward distribution. Then, the gym environment wraps the resultant data. We compose each environment step with one interaction. The provided observation contains the associated user and its metadata, as well as the contextual information from the log. As selected action, the agent should return the recommendation of one partner. The environment also provides additional informative data via a dictionary (for example, a pre-selected list of potential recommendations, in order to narrow the action space).

We compute the reward by comparing the agent's selected action and the partner provided by the log. The agent receives a positive reward if they match. Therefore, the agent only discovers the targeted partner in the scenario of a successful recommendation. Otherwise, it should explore the actions to build its knowledge.

The sequence of steps follows the sequence of interactions in the filtered ground-truth dataset. Hence, we maintain the same temporal dynamic. We define an episode as one iteration trough all logs, rather than the user session. This behavior intends to approximate the multi-agent scenario and keep the perspective on the marketplace, not solely in the user. Finally, the interactions between the proposed agent and the environment generate new interaction logs. This simulated data trains the agent and also provides the cumulative reward curve as the first source of evaluation. In the next subsection, we describe the design of \textit{MARS-Gym} to accomplish this simulation.

\subsection{System Design}

We compose \textit{MARS-Gym} with three main internal components: The Data Engineering Module, the Simulation Module, and the Evaluation Module. Figure \ref{fig:gym} shows a visual representation of our implementation. 
To put them all together, we use Luigi \cite{bernhardsson2020luigi}, which is a Python library for workflow management. This library allows the creation of a graph of tasks, where the output of one task is the input of the next. We used it to compose the graph of tasks from the data processing to the simulation and evaluation. Luigi makes sure that each task (with a set of parameters) only runs once, which is very handy to avoid running the same task several times. In the remainder of this subsection, we describe each module separately.

\begin{figure}[ht]
  \centering
  \includegraphics[width=1\linewidth]{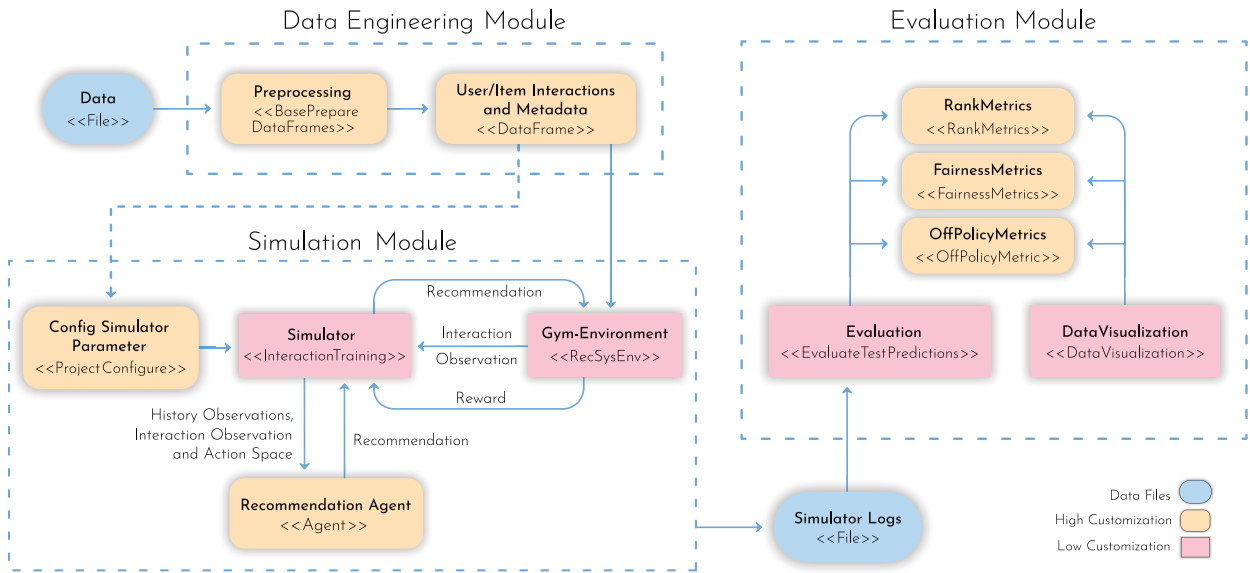}
  \caption{\textit{MARS-Gym} Architecture and its three internal modules.}
  \label{fig:architecture}
\end{figure}



\subsubsection{\textbf{Data Engineering Module}}\label{subsubsection:dataengineering}

This module processes the incoming data to satisfy all requirements to create the OpenAI's Gym environment. It cleans the dataset and applies all transformations to provide the list of interactions and metadata associated with the user or partners as output.

We ingest the dataset as Pandas DataFrames \cite{reback2020pandas,mckinney2010pandas}. Each row should have the user identifier, the fields containing contextual information, the action (partner) presented to the user, and the success flag (if an interaction is successful in the marketplace scenario). All the fields must be in numerical, categorical or NumPy array \cite{oliphant2006numpy} format. We put this limitation in place to be compliant with OpenAI API, but we can easily solve it through the preprocessing steps of the dataset.

To process the dataset, we use Apache Spark \cite{matei2016spark}, which is a powerful distributed computing engine for big data processing with Luigi support. It allows us to have a much higher performance even when running on a single computer. We found that this framework considerably improves the performance to process large datasets inside \textit{MARS-Gym}.

Naturally, distinct marketplaces collect data in many different ways. Therefore, it is impossible to provide a single data processing task that handles all scenarios. For this reason, this module is highly customizable, and we provide abstractions and tools to help in creating data processing tasks. 

\subsubsection{\textbf{Simulation Module}}\label{subsubsection:simulationmodule}

This module is the core of \textit{MARS-Gym} and implements the MDP simulation described in Subsection \ref{sec:mdp_simulation}. Additionally, it is also the module to design and train learning agents. We compose \textit{MARS-Gym} with three main components: the Recommendation Agent, the Gym Environment, and the Simulator.

As previously described, the Gym Environment implements the interface of OpenAI's Gym. It ingests the processed dataset to create a representation of marketplace dynamics by using the same successful interactions and computing the rewards using their corresponding actions as ground truth. On the other side, the Recommendation Agent module implements the agent's interface, which exposes the \textit{act} and the \textit{fit} methods. Specifically, \textit{MARS-Gym} requires the \textit{act} method to return not only the action but also the respective probabilities of each available action. Therefore, any decision-making system that satisfies these requirements can run a simulation.

For the \textit{fit} method, \textit{MARS-Gym} expects to train the agent using previous experience. We already provide high-level PyTorch modules with parameterized architectures (including logistic regression, factorization machine, and neural networks), state-of-the-art gradient-based optimization algorithms, and regularization methods. As we also implement the whole PyTorch's optimization procedure (with GPU support), the only work we left for the user is to combine these building blocks and select the appropriated hyperparameters.

The third component is the Simulator, which plays the central role for the module. It manages both agent and environment to conduct the simulation across episodes, accumulating the experience data for training and the rewards for evaluation. The proposed architecture assumes low customization in this module, as well as for the Gym Environment. In other words, we do not expect any modification in these modules in order to simulate new agents or marketplaces.

The Simulator is agnostic to the nature of training, supporting both online and batch learning, which is easily configurable. However, we choose to do batch learning in our baseline implementations to be closer to what looks like a production system. In such systems, it is often infeasible to learn online with each interaction. So, it is common to retrain the model with a given schedule (e.g., every day or week). To support this kind of off-policy learning, we apply the loss correction via counterfactual estimators, as described in Equation \ref{eq:crm}.

\subsubsection{\textbf{Evaluation Module}}\label{sec:evaluation-protocol}

\begin{figure}[ht]
    \centering
    \includegraphics[width=0.33\textwidth]{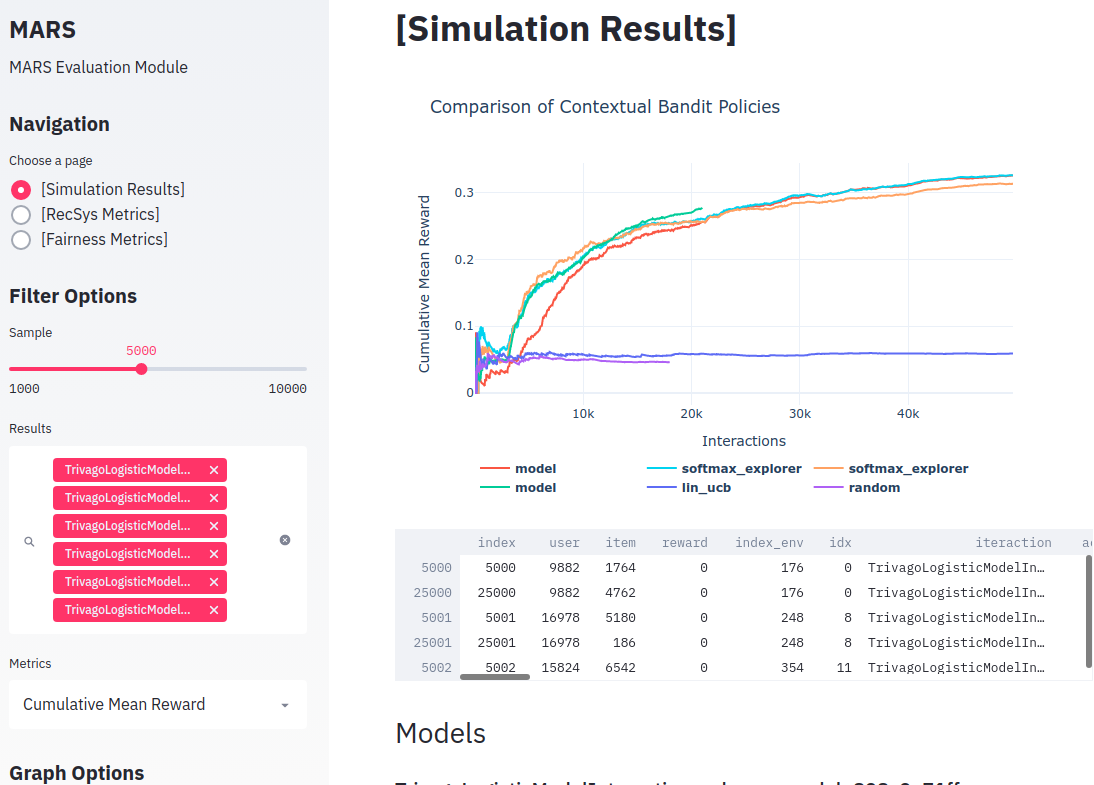} 
    \includegraphics[width=0.33\textwidth]{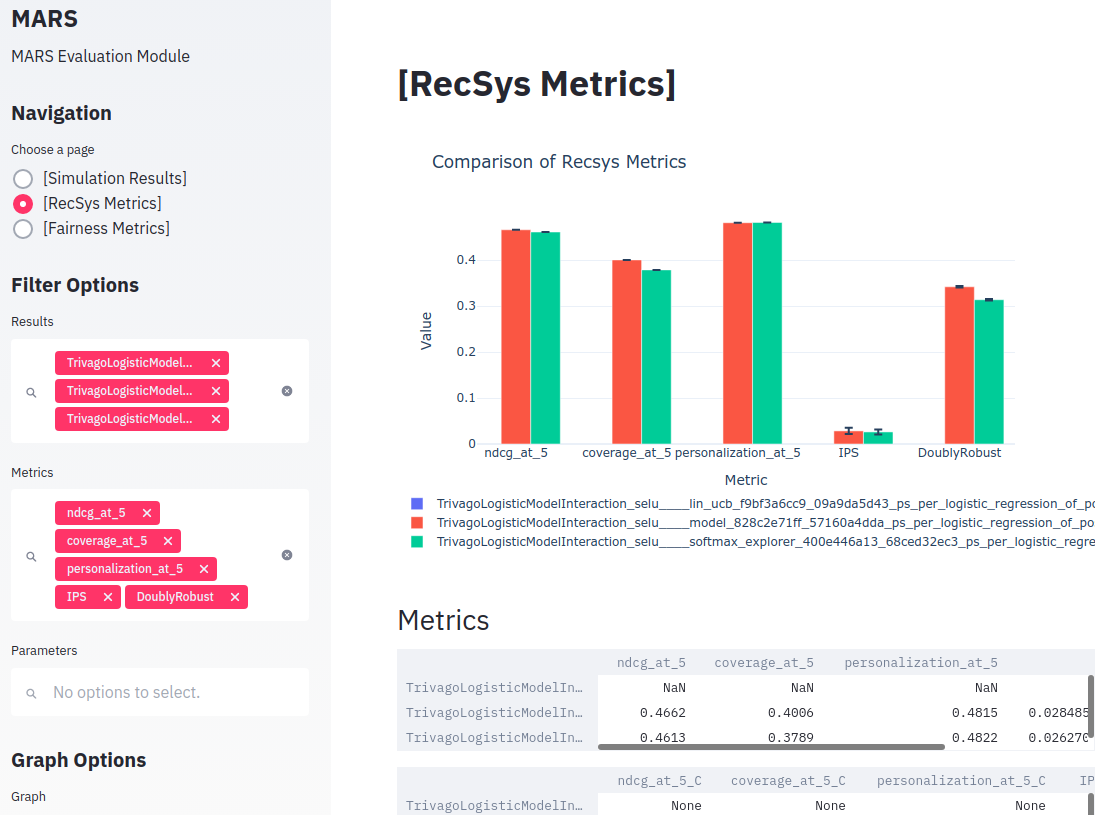}
    \includegraphics[width=0.33\textwidth]{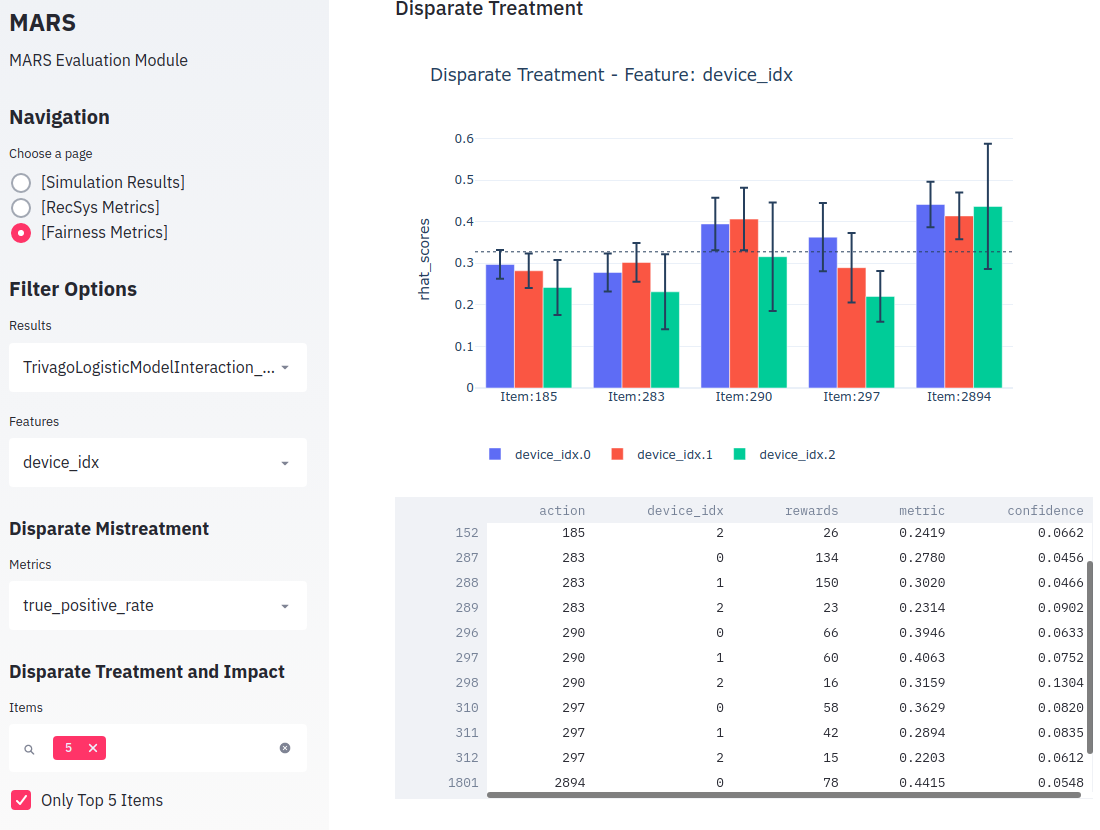}
    \caption{Visualization: \textit{MARS-Gym} provides a user-friendly interface to facilitate feature analysis and the comparison among learning agents.}
  \label{fig:evaluation-visualizations}
\end{figure}

This final module inputs the logs generated from MDP simulation to perform a multi-sided evaluation. The evaluation task computes metrics of recommendation, off-policy evaluation, and fairness. Then, the data visualization component provides a user-friendly interface to present and compare these metrics among different agents, as presented in Figure \ref{fig:evaluation-visualizations}.

The Simulation Module generates two different types of logs to be ingested by the Evaluation Module. During the simulation, it splits the processed dataset into train and test subsets. Naturally, it uses the training subset to conduct the training process, and the Evaluation Module uses the resultant logs to compute and plot the cumulative mean reward. These curves evaluate agent's adaptability, asymptotic performance, and sample efficiency, which are crucial factors to estimate how costly it is to deployment in production. On the other hand, the test subset provides a new perspective of the same MDP, but the agent interacts with it only \textit{after} training. In this way, we use it to measure generalization, as inputs for the evaluation task.

These evaluation logs save the necessary information to compute all metrics. For traditional, rank metrics for Recommender Systems, we use relevance lists to compute Precision, Mean Average Precision \cite{10.1145/1277741.1277790}, and nDCG \cite{10.1145/345508.345545}, as well as the ranked list of actions to compute Coverage \cite{10.1145/1864708.1864761} and Personalization. For this purpose, we can only use the logs that come from successful interactions in the test subset.




In terms of off-policy evaluation, we implement the policy value estimators listed in Subsection \ref{sec:off-policy}. 
 They are especially important to compare against traditional metrics, once they apply corrections to handle the bias originated from collection policy. For example, if an agent \textit{A} has similar nDCG but low Doubly Robust estimation than agent \textit{B}, it means that \textit{A} has biased more towards its collection policy than \textit{B}.

For these metrics, we first need to compute the expected reward estimator, $\hat{\varrho}(x, a)$, and the collection policy estimator, $\hat{\pi_{c}}$, as described in Subsection \ref{sec:off-policy}. They are automatically trained using the whole processed dataset (i.e., the output of the Data Engineering Module) and the same base model of the agent. Then, we use the ground-truth actions and the list of probabilities to compute the metrics.

Finally, for fairness metrics, we propose implementations for equations \ref{eq:treatment}, \ref{eq:impact}, and \ref{eq:mistreatment}, to analyze the notions of disparate treatment, impact, and mistreatment, respectively. To analyze disparate treatment, we compare the policy distribution over all actions and the same distribution conditioned on the sensitive attribute. For that matter, we marginalize over interactions, rather than consider them separately, which we considered infeasible in large scale.

In terms of disparate impact, our implementation chooses a limited set of popular actions and, for each of them, plots the scores associated with each possible value of a sensitive attribute. In this way, we can compare how the attribute values impact an action differently. Finally, for mistreatment, we consider the ground-truth actions in interactions logs to compute classification metrics (such as accuracy, true positive rate, and others), grouped by the values of the sensitive attribute. To illustrate our implementations, we refer to Subsection \ref{sec:baselines}.

%% file: tex/experiments.tex
\section{Experimental Results and Analysis}\label{sec:experiments}

In this section, we describe the Trivago marketplace, which we ingested into \textit{MARS-Gym} for experiments. Furthermore, we present our baseline agents and implementation details, their evaluation results, and share our insights and interpretations about the metrics and how they should impact the marketplace. We hope to illustrate the usage of the framework and, therefore, to facilitate the work of new users.

\subsection{Trivago Marketplace}

Trivago is a global hotel search platform located in more than 190 countries and provides access to more than two million hotels for travelers. 
It centralizes hotel offers on a single platform. Hence, the platform needs to give good recommendations to users and to ensure that affiliated hotels are treated fairly by the platform in different scenarios. Furthermore, user preferences change over time and depend on the trip purpose, as well as accommodation prices, type, and availability. Finally, there is an interest of all partners (traveler, advertising booking site, and Trivago) to suggest good recommendations in different aspects  \cite{RecSysCh73:online}.

\subsubsection{Dataset Description and Benchmark Tasks}

Trivago organized the ACM RecSys Challenge in 2019 \cite{ACMRecSy58:online}. For this competition, it provided a dataset that consists of session logs with 910k samples. Each session contains a sequence of interactions between a user and the platform. They can represent different actions, such as rating, get item metadata (info, image, and deals), sort list, search for a destination or point of interest. In addition to the user session information, the dataset also provides different item metadata that characterize the hotels. Table~\ref{tab:dataset_stats_per_city} shows the general statistics of the Trivago challenge dataset.


We organized the dataset in five tasks, which are primarily identified by the city of interest. In Table \ref{tab:dataset_stats_per_city}, we present the statistics for each proposed task. The task ``RecSys Cities" contains a set of 12 cities presented in the dataset. We chose cities based on several factors: the \textbf{episode length}, to represent how long is the sequence of actions that the agent handles; the \textbf{state space size}, represented by the number of unique sessions ans users, which exposes the variability of contextual information; the \textbf{action space size} (i.e., the number of available partners for recommendation), which indicates how complex is the exploration problem; and the \textbf{number of clickouts}, which gives an idea of how many successful interactions the city has to be used as ground-truth actions during the simulation. For further information, we refer to the dataset page \cite{ACMRecSy58:online} and our source code.

\begin{table}
    \caption{Trivago marketplace -- dataset statistics and benchmark tasks}
    \label{tab:dataset_stats_per_city}
    \begin{tabular}{lrrrrrrc}
    \toprule
         Task & Cities & Interactions &   clickouts & user\_id\_unq & session\_id\_unq & item\_id\_unq \\ 
    \midrule    
        Trivago RecSys Challenge & 34,752 & 15,932,992 & 1,586,586 & 730,803 & 910,683 & 853,540 \\ 
        \hline
        RecSys Cities           & 12        & 761,702    & 62,168     & 31,075  & 36,846  & 12,002   \\
        New York, USA           & 1         & 223,320    & 18,160     & 9,158   & 10,869  & 1,961    \\ 
        Rio de Janeiro, Brazil  & 1         & 161,973    & 9,122      & 4,429   & 5,563   & 2,080    \\ 
        Chicago, USA            & 1         & 22,939     & 1,890      & 1,155   & 1,347   & 662     \\ 
        Como, Italy             & 1         & 1,718      & 155       & 112    & 122    & 328     \\ 
  \end{tabular}
\end{table}

\subsection{Contextual Bandits Baselines}\label{sec:baselines}

We implemented a diverse set of Contextual Bandits as baselines for easy extension and comparison with other agents. They apply a variety of explorations strategies, most of them inspired by \citeauthor{DBLP:journals/corr/abs-1811-04383} \cite{DBLP:journals/corr/abs-1811-04383}. We also proposed CustomLinUCB, which extends LinUCB by implementing a non-linear reward model.

An agent consists of an oracle that predicts scores for each possible interaction between the user and the available actions in the specific context. Therefore, the exploration strategies use these scores to compute probabilities and choose an action. We trained the agents via off-policy batch learning, applying the CRM loss function. The training procedure happened in epochs of a fixed number of interactions, using the whole experience buffer acquired throughout the simulation. Finally, we also performed hyperparameter tuning in all agents and presented the best set found for each of them.

\subsubsection{Bandit Simulation Results}

\begin{figure}[h]
    \centering
    \begin{subfigure}[b]{0.33\textwidth}
        \includegraphics[width=\textwidth]{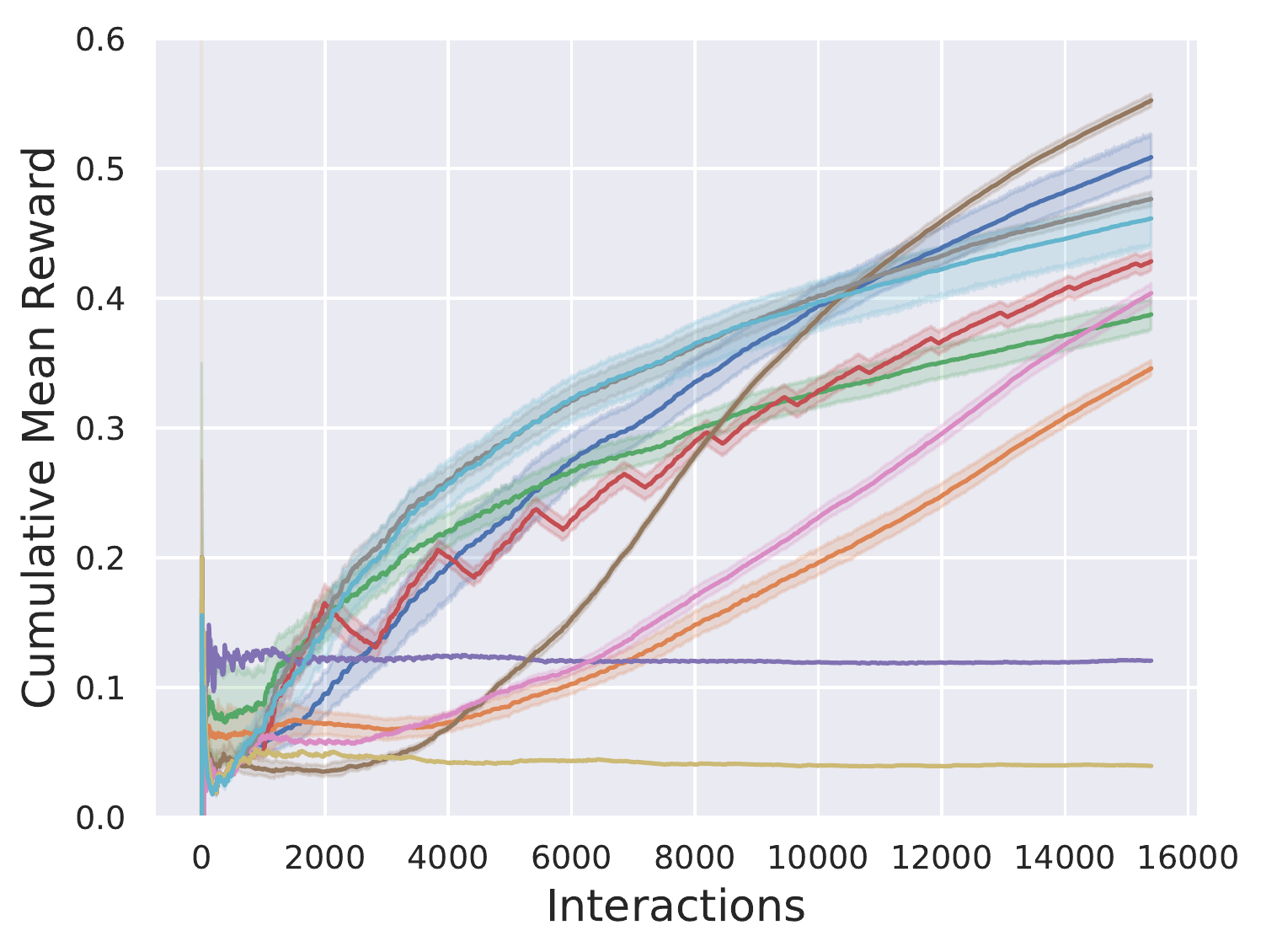}
        \caption{Como, Italy}
        \label{fig:cumulative_mean_reward_italy}
    \end{subfigure}
    \begin{subfigure}[b]{0.33\textwidth}
        \includegraphics[width=\textwidth]{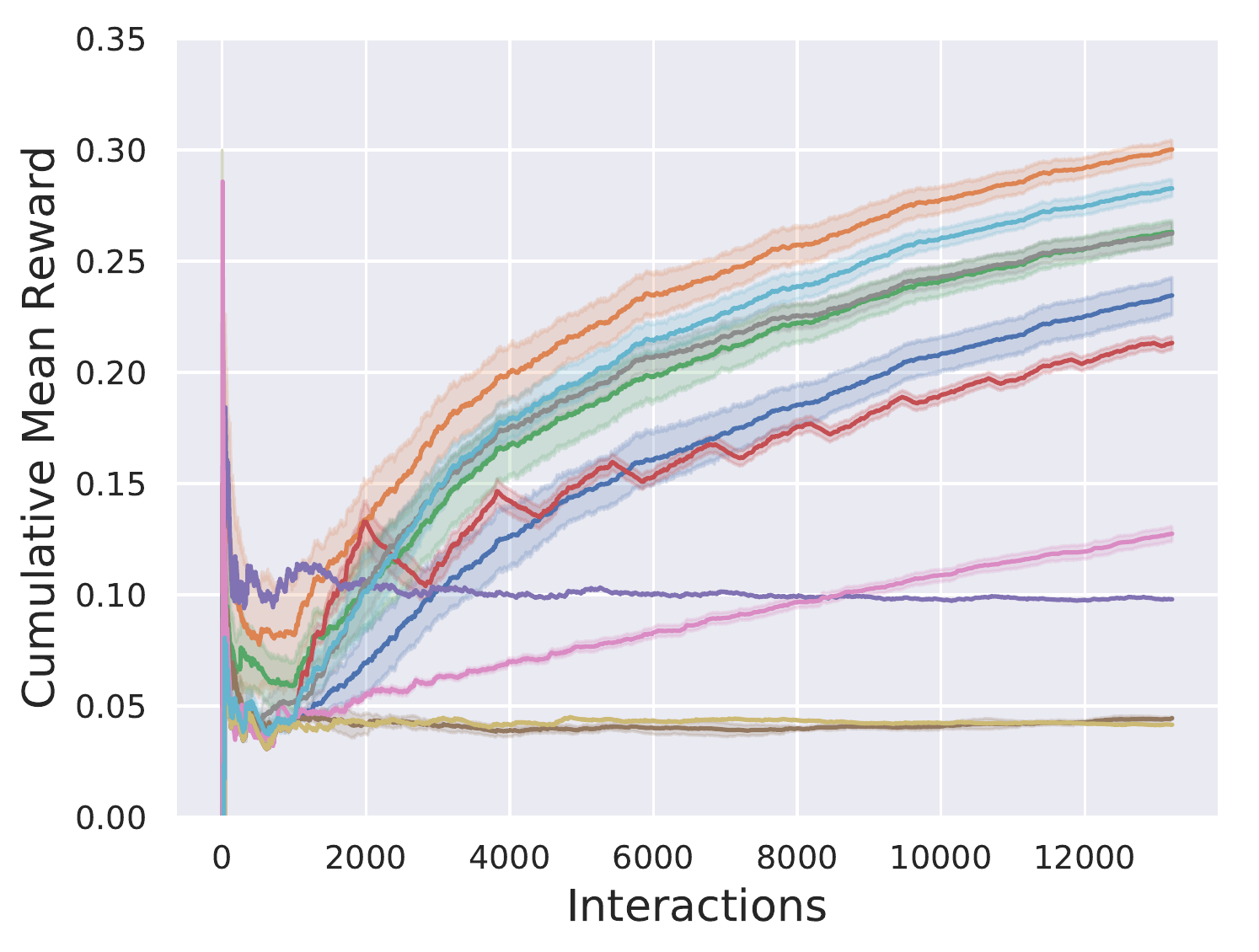}
        \caption{Chicago, USA}
        \label{fig:cumulative_mean_reward_chicago}
    \end{subfigure}
    \begin{subfigure}[b]{0.33\textwidth}
        \includegraphics[width=\textwidth]{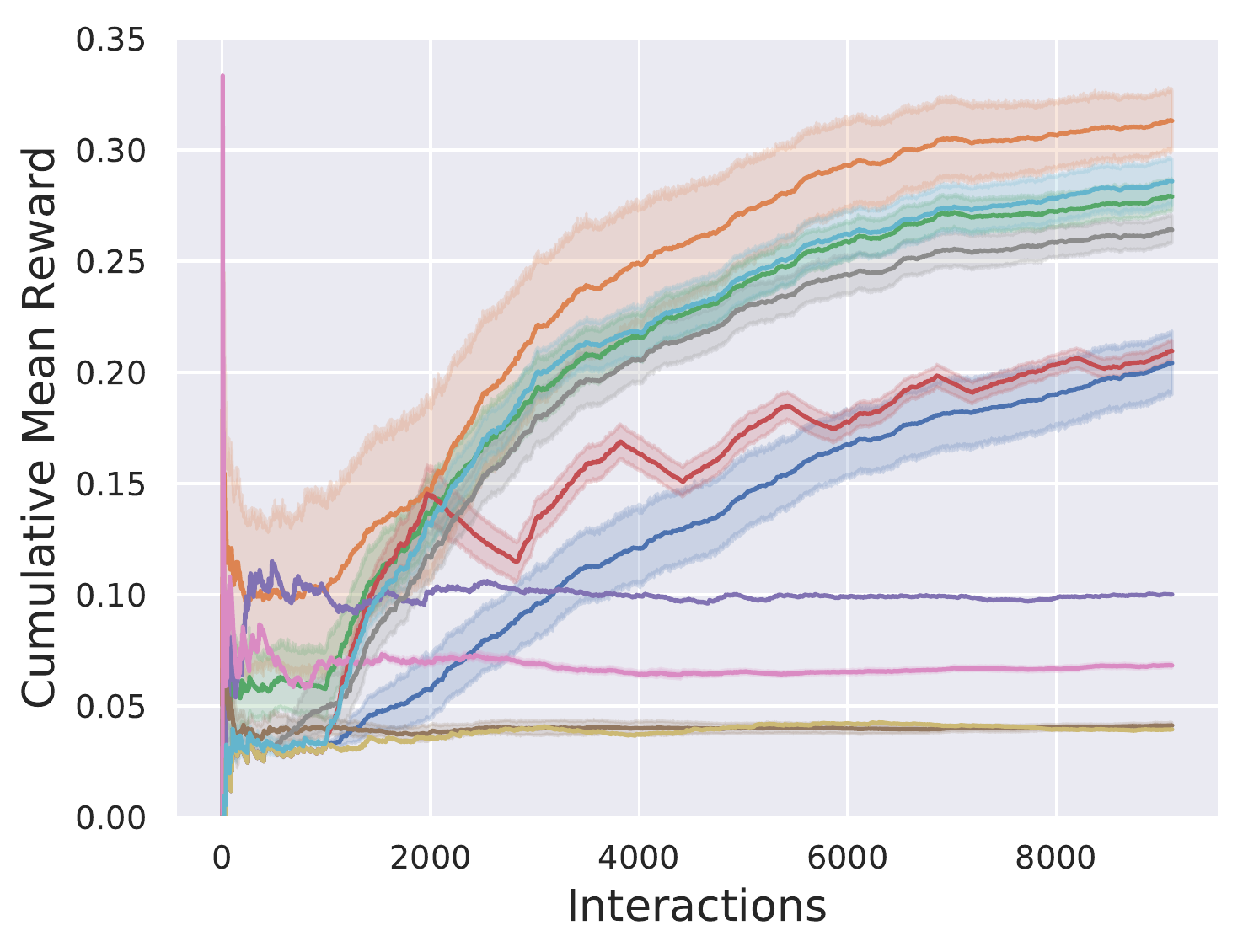}
        \caption{Rio de Janeiro, Brazil}
        \label{fig:cumulative_mean_reward_rio}
    \end{subfigure}
    \begin{subfigure}[b]{0.33\textwidth}
        \includegraphics[width=\textwidth]{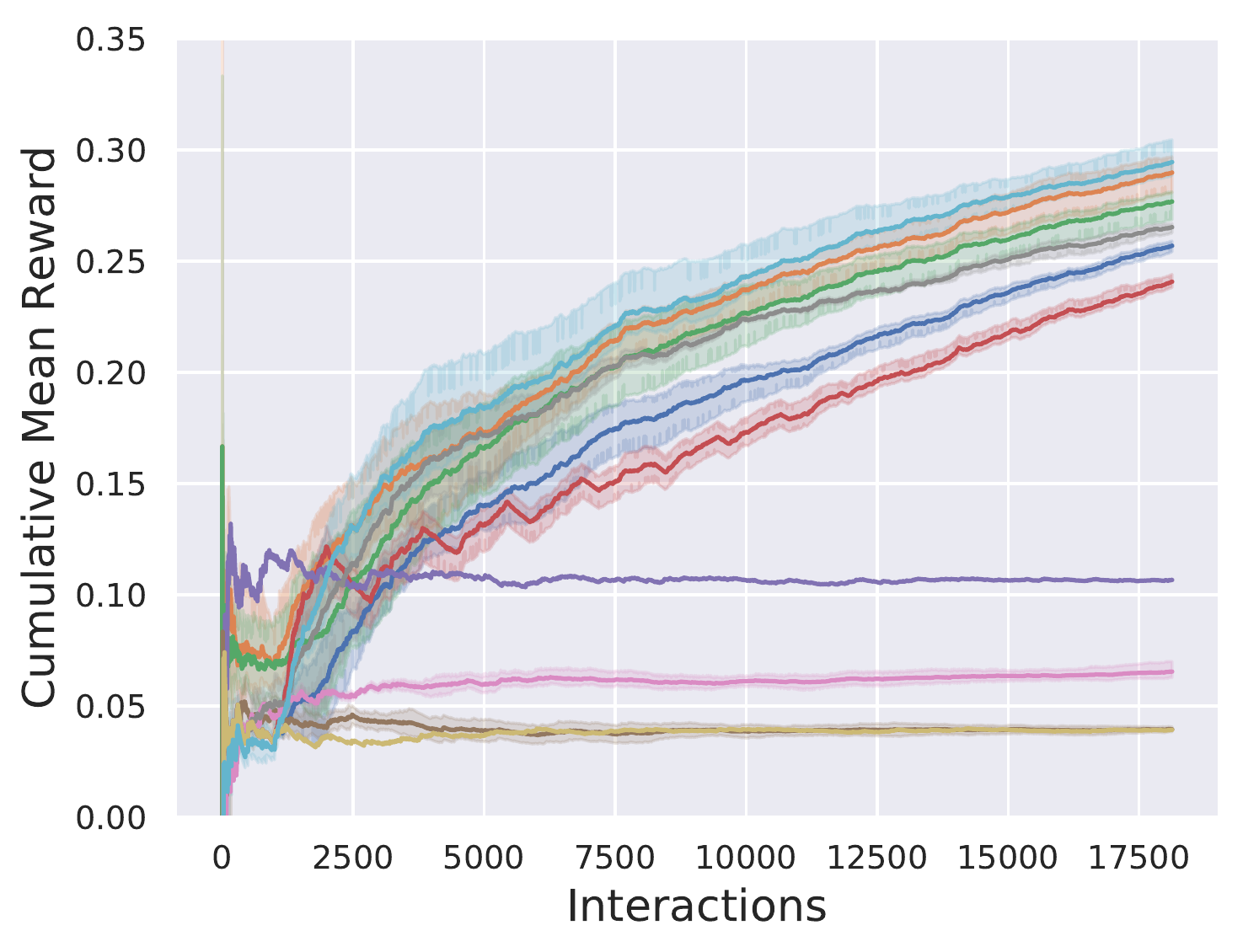}
        \caption{New York, USA}
        \label{fig:cumulative_mean_reward_ny}
    \end{subfigure}  
    \begin{subfigure}[b]{0.45\textwidth}
        \includegraphics[width=\textwidth]{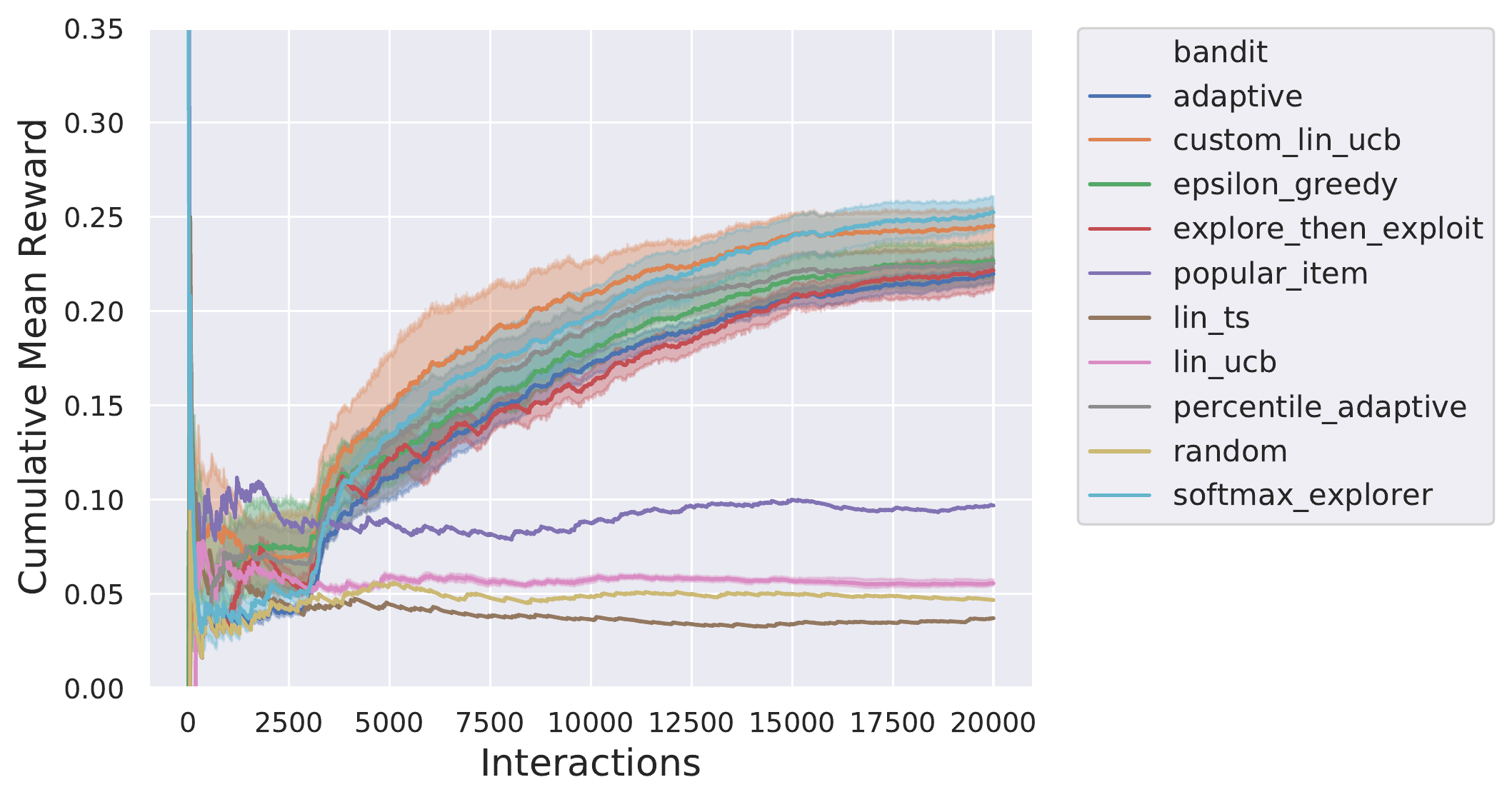}
        \caption{RecSys Cities}
        \label{fig:cumulative_mean_reward_recsys}
    \end{subfigure}      
    \caption{Bandit Simulation Results.}
    \label{fig:cumulative_mean_reward}
\end{figure}

We performed the simulation of each proposed task until the convergence of most methods to observe cumulative mean reward throughout the simulation. It is essential to understand the behavior of the contextual bandits and to make a comparison among them. For statistical significance,  we represent each curve by the mean and confidence interval across five executions of the same experiment. Figure \ref{fig:cumulative_mean_reward} presents the results of the simulations for different tasks and methods.

In general, the methods start with an average reward below the Most Popular agent. As they explore new recommendations and contexts, each method presents a different learning path, but the majority of them surpass this baseline. Furthermore, the performance of these methods dramatically varies across tasks, which reinforces that there is no better exploration strategy for all scenarios. In fact, their performance intrinsically relates to the MDP constraints.

Nevertheless, the experiments also show some patterns among the tasks. We observe that linear methods like LinTS and LinUCB obtain better results when the search space is smaller but does not scale well as the dimensionality of the MDP increases. On the other hand, the proposed CustomLinUCB presents better results in these scenarios but has worse sample efficiency in the low data regime. We hypothesize that the introduction of non-linearities in the oracle increases the capacity of the representation of contextual information in the policy, at the cost of a harder optimization problem.

The curve of the ExploreThenExploit agent, which locally interleaves peaks and valleys as the method switches from full exploitation to full exploration are important to highlight. It suggests a sensibility of the switch hyperparameter, which is undesirable for production environments. Ultimately, we also point out that simple methods (in an implementation perspective) such as $\epsilon$-greedy and softmax explorer present satisfactory results, suggesting to be ideal for composing more complex agents and algorithms as initial setup.

\subsubsection{Recommendation Metrics and Off-Policy Evaluation}

We evaluated the bandits according to traditional recommendation metrics and off-policy metrics in the test subset of the "Chicago, USA" task. Table \ref{tab:recsys_metrics} presents the average results over five executions.

\begin{table*}[h]
    \caption{Recommendation Metrics for "Chicago, USA" task.}
    \label{tab:recsys_metrics}
    \begin{tabular}{lcccccccc}
    \toprule
        & \multicolumn{4}{c}{Classic Recommendation Metrics} & \multicolumn{4}{c}{Off-Policy Metrics}  \\
          \cmidrule(lr){2-5}\cmidrule(lr){6-9} \\
        & \multicolumn{1}{c}{Precision} & \multicolumn{1}{c}{NDCG@5} & \multicolumn{1}{c}{Cove@5} & \multicolumn{1}{c}{Per@5} & \multicolumn{1}{c}{IPS} &  \multicolumn{1}{c}{SNIPS} &  \multicolumn{1}{c}{DE} & \multicolumn{1}{c}{DR} \\           
    \midrule         
        CustomLinUCB                & \textbf{0.338} & 0.443          & 0.371          & 0.724          & \textbf{0.324} & 0.319            & 0.180 & \textbf{0.299} \\ 
        AdaptiveGreedy              & 0.331          & 0.415          & 0.379          & 0.770          & 0.314          & \textbf{0.331}   & \textbf{0.181} & 0.291 \\ 
        PercentileAdaptiveGreedy          & 0.319          & 0.426          & 0.364          & 0.750          & 0.306          & 0.307            & 0.171 & 0.279 \\ 
        SoftmaxExplorer             & 0.316          & \textbf{0.446} & 0.328          & 0.727          & 0.308          & 0.302            & 0.171 & 0.281 \\
        ExploreThenExploit          & 0.315          & 0.423          & 0.313          & 0.737          & 0.308          & 0.316            & 0.167 & 0.280 \\ 
        $\epsilon$-Greedy           & 0.311	         & 0.441          & 0.343          & 0.737          & 0.305          & 0.322            & 0.165 & 0.278 \\ 
        LinUCB                      & 0.065          & 0.183          & 0.297          & 0.721          & 0.047          & 0.049            & 0.033 & 0.043 \\ 
        LinTS                       & 0.029         & 0.112           & \textbf{0.419} & \textbf{0.778} & 0.031          & 0.033            & 0.018 & 0.031\\ 
        MostPopularItem             & 0.074         & 0.199           & 0.153          & 0.592          & 0.063          & 0.052            & 0.046 & 0.061 \\ 
        Random                      & 0.042         & 0.145           & 0.392          & 0.777          & 0.028          & 0.031            & 0.024 & 0.029 \\ 
        \hline
        Policy without CRM          & \textbf{0.354} & \textbf{0.457} & \textbf{0.303} & 0.717 & 0.340 & 0.346 & \textbf{0.199} & 0.312 \\ 
        Policy with CRM             & 0.344 & 0.434 & 0.300 & \textbf{0.730} & \textbf{0.350} & \textbf{0.352} & 0.195 & \textbf{0.320} \\ 
  \end{tabular}
\end{table*}

Firstly, we confirm the hypothesis that the precision metric highly correlates to the simulations results in Figure \ref{fig:cumulative_mean_reward_chicago}. CustomLinUCB presented best the results in the task, as in the simulation. Comparatively, AdaptiveGreedy and ExploreThenExploit performed better in evaluation, while $\epsilon$-Greedy performed worse than the other methods. We hypothesize that these changes are related to the sample efficiency of each exploration strategy, as well as the generalization properties from the learned policies. Furthermore, we observe that the SoftmaxExplorer outperforms all other methods in the nDCG metric, which suggests good performance in scenarios of slate recommendation. Finally, we highlight the AdaptiveGreedy performance for diversity metrics (Coverage and Personalization), while maintaining good accuracy metrics.

A consistent drop from the precision to the off-policy metrics can be observed, suggesting that all methods naturally exploit the sampling bias. Comparatively, we see a high correlation between those metrics. However, the ranking of bandits is not the same, which suggests that some bandits suffer more from sampling bias. This evidence and analysis are important for model deployment, once that off-policy metrics often correlates better with online experiments \cite{jeunen2019value, jeunen2019learning}. 

Ultimately, we conducted an ablation study regarding CRM loss.  For this purpose, we trained two policies using the training data in a supervised way (i.e., no simulation), varying the loss function. From Table \ref{tab:recsys_metrics}, we observe that the policy with CRM trades off accuracy in recommendation metrics to improve the unbiased, off-policy metrics. This result validates that this technique is essential to reduce the effect of sampling bias during training, and, therefore, improve off-policy evaluation.

\subsubsection{Fairness Results}\label{subsec:fairnessresult}

\begin{figure}[h]
    \centering
    \begin{subfigure}[b]{0.33\textwidth}
        \includegraphics[width=\textwidth]{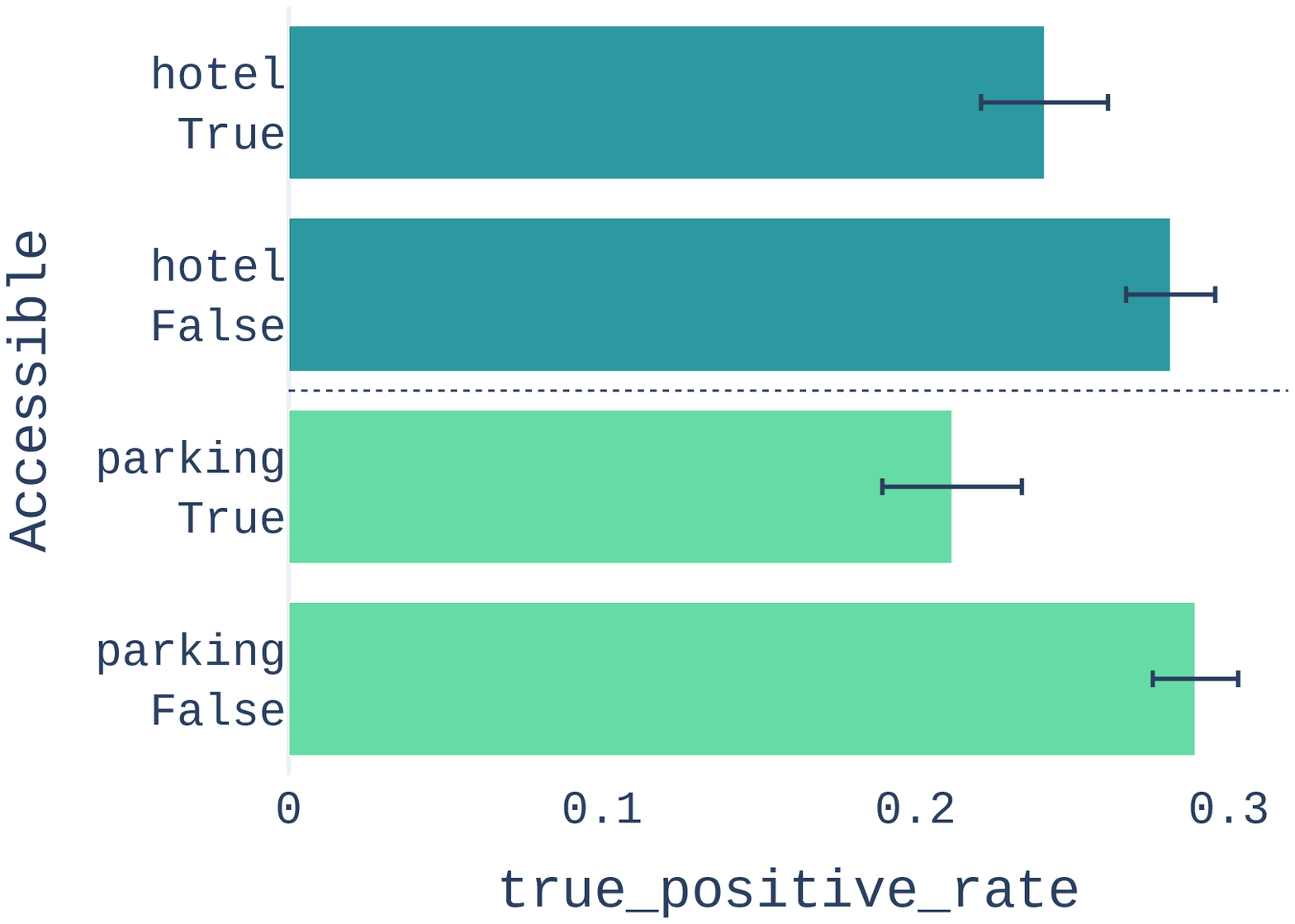}
        \caption{Feature -- Accessibility}
        \label{fig:fairness_accessible}
    \end{subfigure}
    \begin{subfigure}[b]{0.33\textwidth}
        \includegraphics[width=\textwidth]{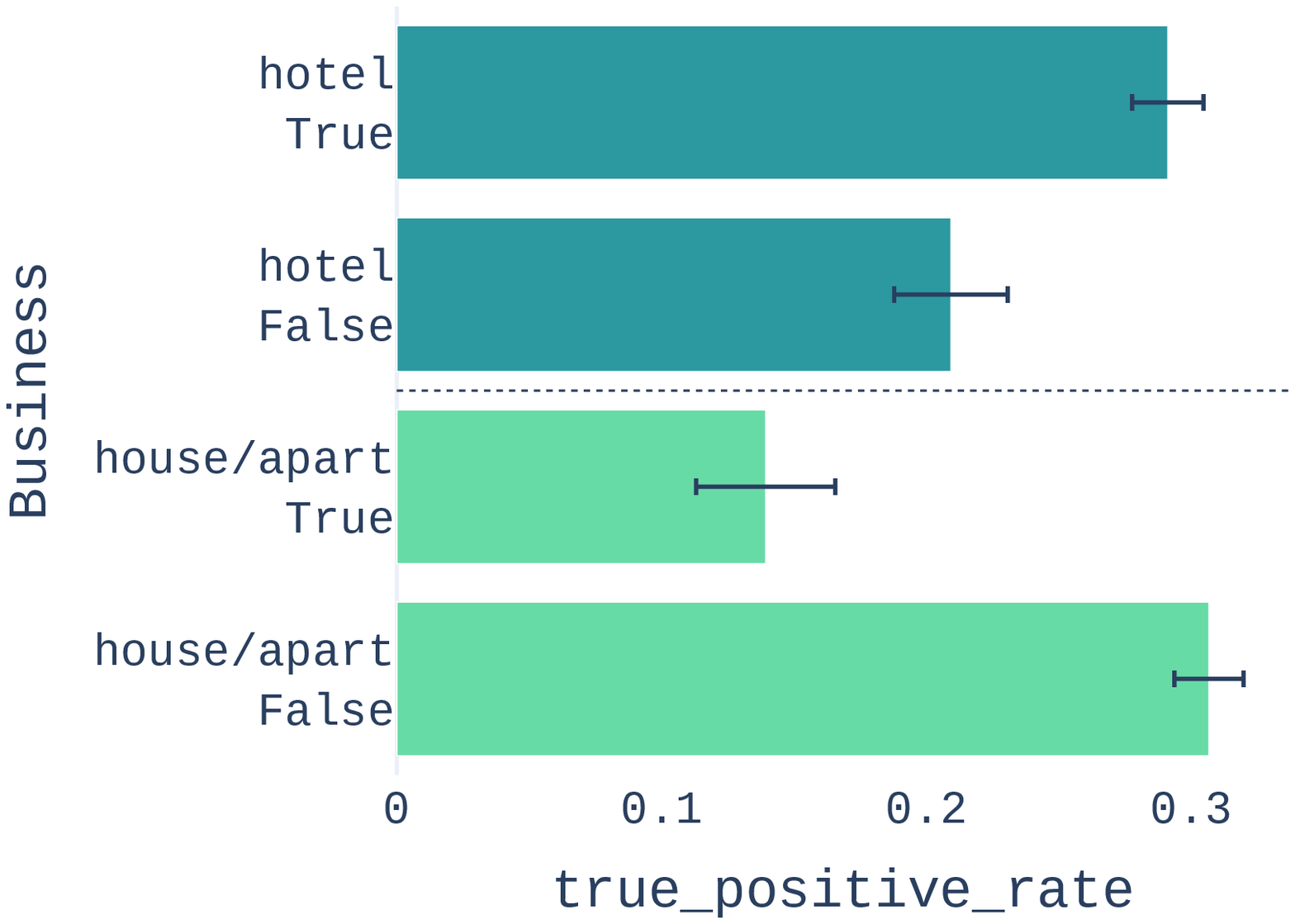}
        \caption{Feature -- Accommodation Type}
        \label{fig:fairness_business}
    \end{subfigure}
    \begin{subfigure}[b]{0.33\textwidth}
        \includegraphics[width=\textwidth]{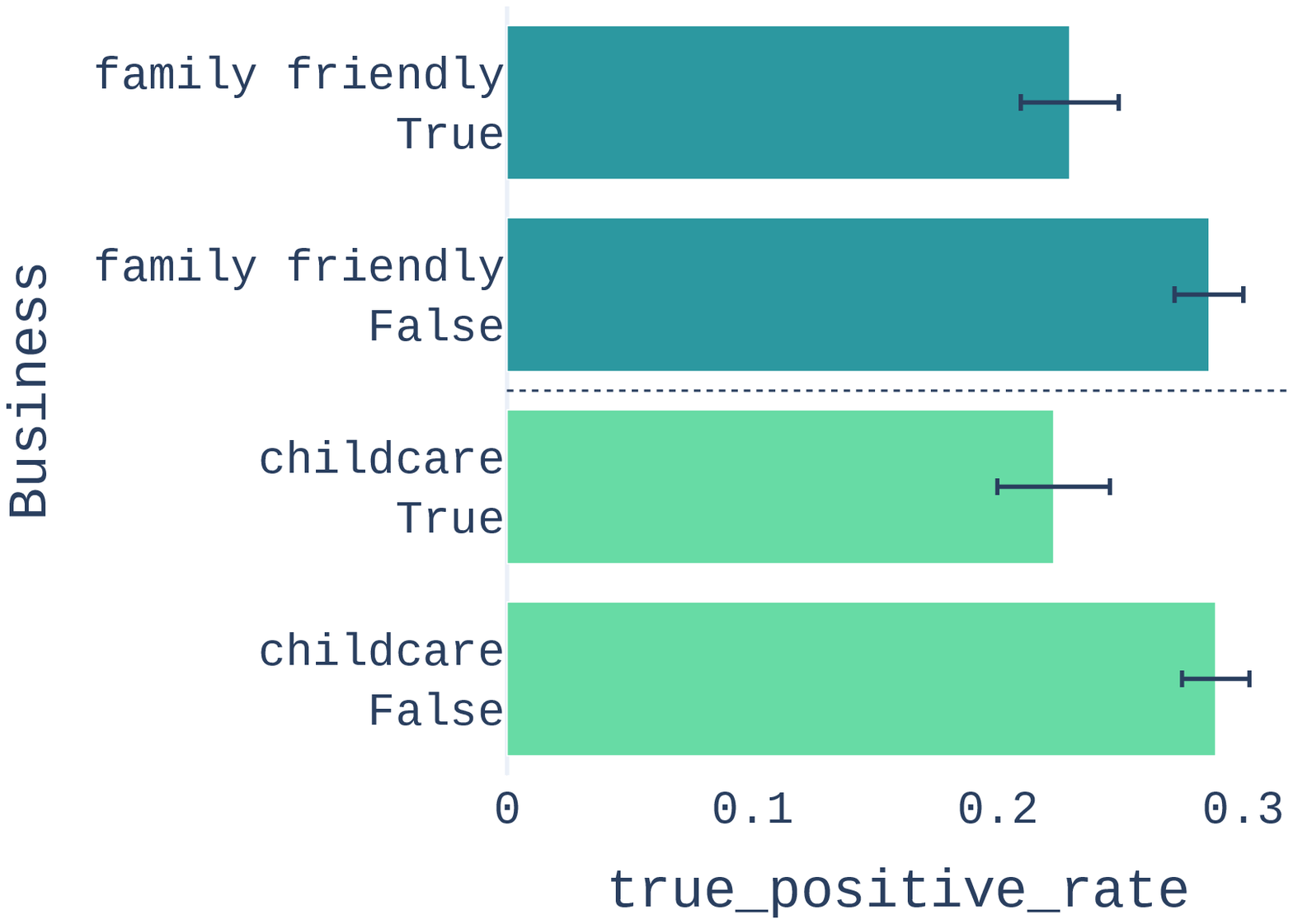}
        \caption{Feature -- Business Affinity}
        \label{fig:fairness_business2}
    \end{subfigure}    
    \begin{subfigure}[b]{0.33\textwidth}
        \includegraphics[width=\textwidth]{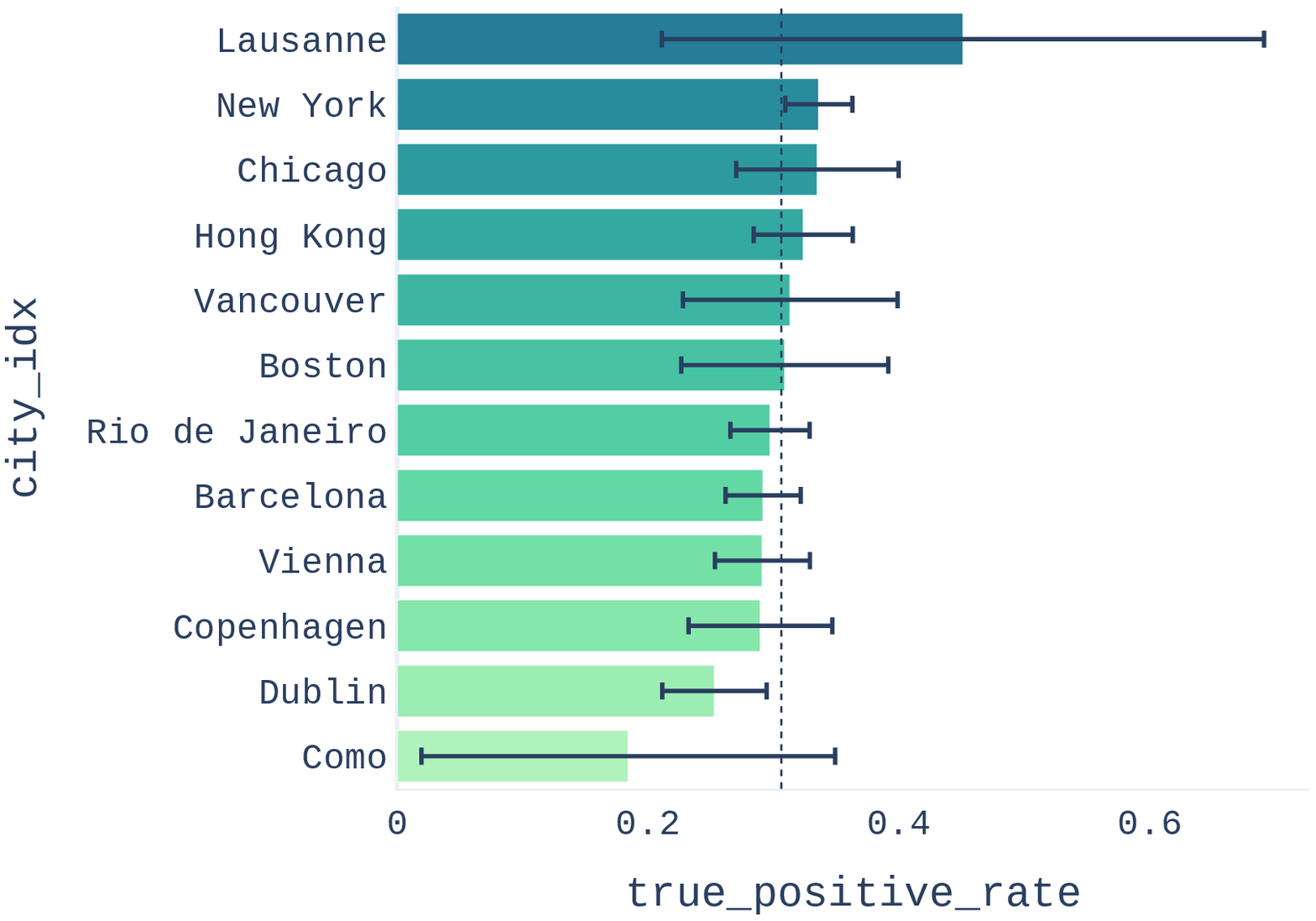}
        \caption{Feature -- City}
        \label{fig:fairness_city}
    \end{subfigure}        
    \begin{subfigure}[b]{0.33\textwidth}
        \includegraphics[width=\textwidth]{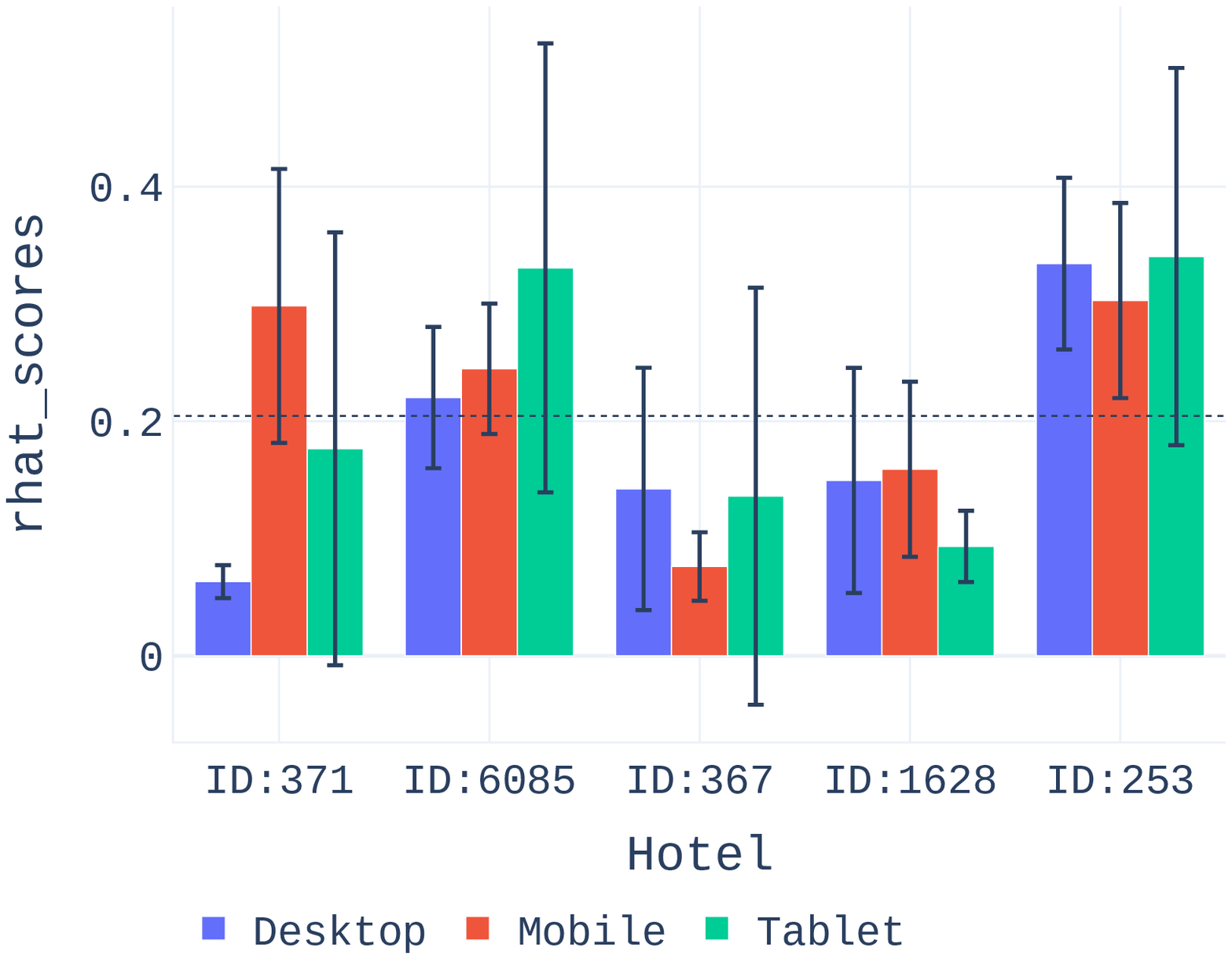}
        \caption{Feature -- Device}
        \label{fig:fairness_device}
    \end{subfigure}    
    
    \caption{Fairness analysis for SoftmaxExplorer.}
    \label{fig:fairness}
\end{figure}

We evaluated the SoftmaxExplorer bandit on the ``RecSys Cities" task, in the perspective of disparate mistreatment (Figures \ref{fig:fairness_accessible}, \ref{fig:fairness_business}, \ref{fig:fairness_business2}, and \ref{fig:fairness_city}) and disparate treatment (Figure \ref{fig:fairness_device}). We selected a few attributes that we judged to be sensitive for all partners in the marketplace.

In Figure \ref{fig:fairness_accessible}, we present true positive rates for two features related to accessibility, which is a requirement for a niche of users. We observe that the bandit provides poorer recommendations for accommodations with accessibility than those without it, suggesting disparate mistreatment, which is undesirable for the experience of users that requires this feature.  Figure \ref{fig:fairness_business}, on the other side, presents features related to the type of accommodation. We observe that hotels receive a much better recommendation from the agent than the other types, such as hostels or apartments. Similarly, Figure \ref{fig:fairness_business2} also presents disparate mistreatment for accommodations that are family-friendly or provides childcare. These are not only business features from partners but also user requirements.

From a different perspective, Figure \ref{fig:fairness_city} shows the true positive rates for each city analyzed. In general, we see similar results for all of them. The metrics for some cities present higher confidence intervals, which is directly related to having fewer data. Nevertheless, we diagnose differences among some cities (e.g., comparing New York and Dublin), which we could explore to understand the source of unfairness and improve the recommender system.

Finally, Fig. \ref{fig:fairness_device} presents the scores from the same bandit for five different hotels grouped by the user's device. Although many of them present the same treatment for all devices, we find that the hotel with ID 371 presents considerably low scores for desktop. We hypothesize that it relates to the user experience using this device. This is a case where disparate treatment arises, and the platform must address it.

Ultimately, all charts in Figure \ref{fig:fairness} diagnose different scenarios in which the recommender system does not satisfy the current notions of fairness. They are the reproduction of many biases in the dataset, and the machine learning pipeline and platform must address it. For example, we can change the user interface for tablet in the platform, or balance the dataset for the different accommodations. Anyway, these metrics provide insights into where the system needs to improve in order to maintain the marketplace fairness and healthiness.

%% file: tex/conclusions.tex
\section{Conclusions and Future Work}\label{sec:conclusions}

In this work, we proposed \textit{MARS-Gym}, an open-source framework to model, train, and evaluate RL-based recommender systems for marketplaces. We presented in detail its internal components and provide baseline implementation and analysis to serve as an introduction for the users. We also point out that the presented results are an extra contribution of our work on the task of benchmarking contextual bandits, complementing the results of previous works. 

As future releases of this framework, we plan to implement extensions for multi-objective  and constrained optimization to extend the support for marketplaces with multiples stakeholders during learning. It addresses, for example, the recommendation in marketplaces with a delivery system, where logistic restrictions arise. We also plan to extend the framework to address the hierarchical setting, where the agent manages multiples recommendation models rather than solely actions. This problem is valuable for current production systems, where we often have a diverse set of models running concurrently and need to evaluate their adaptability and degradability over time.

Finally, we also expect to receive contributions from the community, not only with feature requests and new code but also with new evaluation tasks, competitions, and ingested datasets. In this way, we hope to popularize this project and provide necessary tooling to accelerate research and development of reinforcement learning agents for marketplace recommendation.

%% file: tex/acknowledgments.tex
\section{ACKNOWLEDGMENTS}\label{sec:acknwledgments}

We would like to thank iFood Research for financial support and helpful reviews.